\documentclass[11pt,english]{article}
\usepackage{lmodern}
\usepackage{color}

\usepackage[T1]{fontenc}
\usepackage[utf8x]{inputenc}
\usepackage{amsthm}
\usepackage{amsmath}
\usepackage{amssymb}
\usepackage{esint}
\usepackage{textcomp}
\usepackage{graphicx}
\usepackage{float}

\makeatletter
  \newtheorem{rem}{\protect\remarkname}
  \theoremstyle{plain}
  
  \theoremstyle{plain}
  \newtheorem{prop}{\protect\propositionname}
  \theoremstyle{plain}

\@ifundefined{date}{}{\date{}}
\usepackage{a4wide}

\def\logexpo{\frac{1}{\gamma}\log\Biggl(\mathbb{E}\Biggl[\exp\Biggl(-\gamma\Biggl(}
\def\logexpc{\Biggr)\Biggr)\Biggr]\Biggr)}

\def \E{\mathbb{E}}
\def \P{\mathbb{P}}
\def \R{\mathbb{R}}
\def \Sum{\displaystyle\sum}

\title{Optimal execution of ASR contracts with fixed notional \footnote{This research has been conducted with the support of the Research Initiative ``March\'es financiers \`a haute fr\'equence'', financed by HSBC France, under the aegis of the Europlace Institute of Finance. The
author would like to thank Nicolas Grandchamp des Raux (HSBC France), Charles-Albert Lehalle (CFM), Jiang Pu (Institut Europlace de Finance) and Mathieu Rosenbaum (UPMC) for the discussions we had on the topic.}}
 \author{Olivier {\sc Gu\'eant} \footnote{Universit\'e Paris-Diderot, UFR de
Math\'ematiques, Laboratoire Jacques-Louis Lions, gueant@ljll.univ-paris-diderot.fr}
 }

\makeatother

\usepackage{babel}
  \providecommand{\lemmaname}{Lemma}
  \providecommand{\propositionname}{Proposition}
  \providecommand{\remarkname}{Remark}

\begin{document}
\maketitle
\begin{abstract}
Be it for taking advantage of stock undervaluation or in order to distribute part of their profits to shareholders, firms may buy back their own shares. One of the way they proceed is by including Accelerated Share Repurchases (ASR) as part of their repurchase programs. In this article, we study the pricing and optimal execution strategy of an ASR contract with fixed notional. In such a contract the firm pays a fixed notional $F$ to the bank and receives, in exchange, a number of shares corresponding to the ratio between $F$ and the average stock price over the purchase period, the duration of this period being decided upon by the bank.
From a mathematical point of view, the problem is related to both optimal execution and exotic option pricing.

\vspace{1cm}

 \vspace{5mm}

\end{abstract}

\section{Introduction}

The traditional way for a firm to repurchase its own shares is through open-market repurchase programs (OMR).\footnote{Privately negotiated repurchases and self-tenders represent a few percents of the total amount of repurchases -- see \cite{chemmanur}.} However, as reported in \cite{bargeron}, after they have announced they intend to buy back shares, a substantial number of firms do not commit to their initial plan. Unexpected shocks on prices or on the liquidity of the stock may indeed provide incentives to slow down, postpone, or even cancel repurchase programs.

In order to make a credible commitment to buy back shares, an increasing number of firms enter Accelerated Share Repurchase contracts with investment banks. Accelerated Share Repurchase contracts are of several kinds and we traditionally distinguish between ASR with fixed number of shares and fixed notional ASR.

In the case of an ASR with fixed number of shares ($Q$), the contract is the following:\footnote{We present here the case of a pre-paid ASR with fixed number of shares. In many cases, especially in Europe, the shares are not delivered at inception but at the end of the contract. Although share repurchase is not \emph{accelerated} in this case, the pricing and hedging of the contract is the same, if we ignore funding and interest rate issues.}
\begin{enumerate}
\item At time $t=0$, the bank borrows $Q$ shares from shareholders (ordinarily institutions) and gives the shares to the firm in exchange of a fixed amount $QS_0$ where $S_0$ is the Mark-to-Market (MtM) price of the stock at time $t=0$. The bank then has to progressively buy back $Q$ shares on the market to give back $Q$ shares to the initial shareholders, and go from a short position to a flat position on the stock;
\item  The bank is long an option with payoff $Q(A_\tau - S_0)$ (the firm being short of this option), where $A_t$ is the average price between $0$ and $t$ (in practice the average of closing prices or the average of daily VWAPs) and where $\tau$ is chosen by the bank, among a set of specified dates\footnote{There is usually a latency period of a few weeks, after $t=0$, during which the option cannot be exercised.} $\tau_1, \ldots, \tau_k$.
\end{enumerate}
Hence, the firm eventually pays on average the price $A_\tau$ for each of the $Q$ shares that have been repurchased.

In the case of a fixed notional ASR with notional $F$, the contract is the following:\footnote{As above, we only discuss pre-paid contracts. Post-paid contracts also exist.}
\begin{enumerate}
\item At time $t=0$, the bank borrows $Q$ shares from shareholders -- usually $Q = \alpha \frac{F}{S_0}$, where $\alpha \in [0,1]$ is usually around $80\%$ --, and gives the shares to the firm in exchange of the fixed amount $F$;
\item  An option is embedded in the contract: when the bank exercises this option at date $\tau$ (the date is chosen by the bank, among a set of specified dates $\tau_1, \ldots, \tau_k$), there is a transfer of $\frac{F}{A_\tau} - Q$ shares between the bank and the firm\footnote{$\alpha$ is chosen small enough so that the transfer is almost always from the bank to the firm, \emph{i.e.} $\frac{F}{A_\tau} \ge Q$. In what follows, we implicitly assume that the settlement is from the bank to the firm.}, so that the actual number of shares obtained by the firm is $\frac{F}{A_\tau}$.
\end{enumerate}
Hence, the firm pays $F$ and eventually gets $\frac{F}{A_\tau}$ shares.

As above, the bank will also progressively buy shares on the market to give back $Q$ shares to the initial shareholders.

Pricing these contracts and hedging them are mathematical problems that cannot be solved in a satisfactory way using the classical theory of derivatives pricing. At time $t=0$, the bank has a short position on the stock and it will buy shares on the market over a period of a few weeks to a few months. As a consequence, the problem is at the same time a problem of optimal execution and a problem of option pricing: a cash-settled exotic option with Bermudean exercise dates in the case of an ASR with fixed number of shares and a physically-settled exotic option with Bermudean exercise dates in the case of a fixed notional ASR. Ignoring the interactions between the two problems would lead to a suboptimal strategy and to mispricing. Moreover, since the nominal/notional of these options is large, execution costs must be taken into account and we propose a framework coming from the literature on optimal execution to price and execute optimally these contracts.

The problem of pricing derivative products with large nominal has already been tackled in the literature using the tools of optimal execution. Rogers and Singh \cite{rogerssingh} considered execution costs that are not linear in (proportional to) the volume executed, but rather strictly
convex to account for liquidity effects (as opposed to the literature on pricing with transaction costs -- see for instance \cite{tcleland}). They considered
an objective function that penalizes both execution costs and mean-squared
hedging error at maturity. They obtained, in this close-to-mean-variance
framework, a closed-form approximation for the optimal hedging strategy
of a vanilla option when illiquidity costs are small. Also, Li and Almgren \cite{li}, motivated by saw-tooth
patterns recently observed on several US stocks (see \cite{lehalle1,lehalle2}),
considered a model with both permanent and temporary impact. In their
model, they assumed execution costs are quadratic and they considered a constant
$\Gamma$ approximation. Using another objective function, they obtained a closed
form expression for the hedging strategy of a call option. Finally, Gu\'eant and Pu proposed in \cite{gueantpu} a method to
price and hedge a vanilla option in a utility-based framework, under
general assumptions on market impact. In particular, they considered both the case of physical
settlement and the case of cash settlement.

The case of ASR contracts is however more complex than the case of vanilla options, because the payoff is not a classical European payoff, but rather an exotic one with Asian and Bermudean features, and because the problem to be solved is both a problem of option pricing/hedging and a problem of optimal execution.

Economists have discussed the role of ASR contracts among the different ways to buy back shares (see for instance \cite{bargeron,chemmanur}). The mathematical literature on Accelerated Share Repurchase contracts is however rather limited and very recent. The first paper on the topic is an interesting paper by Jaimungal, Kinzebulatov
and Rubisov \cite{jaimungal}. The authors focus on ASR contracts with fixed number of shares. For finding the optimal buying strategy, they propose a model in continuous time with the following characteristics: stock prices are modeled with a perturbed geometric Brownian motion (the perturbation accounts for permanent market impact), execution costs are quadratic as in the original Almgren-Chriss models \cite{almgren,almgren2}, the range of possible exercise dates is $[0,T]$ (hence the product is American rather than Bermudean). The optimal strategy is found in \cite{jaimungal} for an agent who is risk-neutral, although the authors add inventory penalties as in \cite{cj1,cj2}. The main interest of  \cite{jaimungal} is that the authors manage to reduce the problem to a 3-variable PDE, whereas the initial problem is in dimension 5. In particular, they show that the exercise boundary only depends on the time to maturity and the ratio between the stock price and its average value since inception. The case of ASR contracts with fixed number of shares has also been considered in a different model by Gu\'eant, Pu and Royer (see \cite{asr}). In this discrete-time model, a risk-averse agent is considered in an expected utility framework. Permanent market impact is considered as in \cite{jaimungal}, but stock prices are assumed to be normal rather than log-normal. Also, a general form of execution cost is allowed as in \cite{gueant}, and exercise dates are among a finite set of dates. As in \cite{jaimungal}, the model in \cite{asr} boils down to a set of equations with 3 variables: time, the number of shares to be bought, and the difference between the current stock price and the average price since time $t=0$. An original fast tree-based numerical method\footnote{Trees are not recombinant but the number of nodes is a cubic function of the number of time periods.} is proposed in \cite{asr}.

The model we propose in this paper is inspired from \cite{asr}. In particular, we use the same expected utility framework. However since we focus on the specific case of fixed notional ASR contracts, the reduction of the problem to a 3-dimension one is not possible anymore. In particular, the original tree-based method proposed in \cite{asr} to approximate numerically the optimal strategy cannot be used. We present in Section 2 the framework of the model without permanent market impact. As in \cite{asr}, the model is in discrete time and we end up with a characterization of the optimal buying/hedging strategy with recursive Bellman equations. In Section 3, we introduce these Bellman equations, along with a change of variables to reduce the dimensionality of the problem (from 5 to 4). We also introduce the indifference price of the ASR contract. In Section 4, we present a numerical method involving trees and splines to approximate the solution of the problem, along with examples. The introduction of permanent market impact is presented in Appendix~A.

\section{Setup of the model and Bellman equations}
\label{sect: setup}
We consider the pricing and hedging of a fixed notional ASR from the point of view of an investment bank buying back shares for a client (a firm).\footnote{The mechanism of the contract is explained in the previous section.} Throughout the paper, the notional of the ASR contract will be denoted by $F$ and the maturity of the contract will be denoted by $T$.

The model we consider is a discrete-time model where each period of time (of length $\delta t$) corresponds
to one day. In other words, if the interval $[0,T]$ corresponds to $N$ days, we assume that decisions are made at times $t_0 = 0 < \ldots < t_n = n \delta t < \ldots < t_N = N \delta t = T$.

In order to introduce random variables, we introduce a probability space $\left(\Omega,\left(\mathcal{F}_{n}\right)_{0\le n\le N},\P\right)$, where the filtration satisfies the usual assumptions. All random variables will be defined on this probability space.

As far as prices are concerned, we start with an initial price $S_{0}$ and we consider
that the dynamics of the price\footnote{Linear permanent market impact as in \cite{gatheral} can be introduced in the model -- see Appendix A. However, we believe that, in the case of ASR contracts, there is little information in trades, and subsequently almost no permanent market impact, since buy-back programs are usually announced before they actually occur.} is given
by:
$$
\forall n\in\lbrace0,\ldots,N-1\rbrace,\quad S_{n+1}=S_{n}+\sigma\sqrt{\delta t}\epsilon_{n+1},
$$
where $\epsilon_{n+1}$ is $\mathcal{F}_{n+1}$-measurable and where the random variables
$\left(\epsilon_{n}\right)_{n}$ are assumed to be i.i.d. with
mean $0$, variance $1$. We also assume that these variables have a moment-generating function defined on $\R_+$.

Depending on the exact payoff of the considered Accelerated Share Repurchase
contract, $S_{n}$ can be regarded as the daily VWAP over the
period $[t_{n-1},t_n]$ ($n\ge 1$), or the closing price of the $n^{\text{th}}$ day.\footnote{The value of $\sigma$ depends on what $S$ stands for.}

The average price process entering the definition of the payoff in the ASR contract is denoted by $\left(A_{n}\right){}_{n\ge1}$:
$$
A_{n} =\frac{1}{n}\sum_{k=1}^{n}S_{k}.
$$

in order to buy shares, we assume that the bank sends every day an order to
be executed over the day. At time $t_n$, the size of the order
sent by the bank is denoted by $v_{n}\delta t$ where the process $(v_n)_n$
is assumed to be adapted. Hence, the number of shares bought on the market to give them back to the initial shareholders is given by:
\[
\left\{ \begin{array}{lcl}
q_{0} & = & 0\\
q_{n+1} & = & q_{n}+v_{n}\delta t.
\end{array}\right.
\]
The price paid by the bank for the shares bought over $[t_n,t_{n+1}]$
is assumed to be $S_{n+1}$ plus execution costs;\footnote{If $S$ models VWAP, we assume implicitly that the bank has sent a VWAP order. If $S$ models closing prices, we implicitly assume that a target close order has been sent by the bank.} the execution  costs being modeled by a function $L\in C(\mathbb{R},\mathbb{R}_{+})$ satisfying the following assumptions:
\begin{itemize}
\item $L(0)=0$,
\item $L$ strictly convex, even and increasing on $\R_{+}$,
\item $L$ is asymptotically superlinear, \emph{i.e.}:
\begin{eqnarray*}
\lim_{\rho\to+\infty}\frac{L(\rho)}{\rho} & = & +\infty.
\end{eqnarray*}
\end{itemize}

The cash spent by the bank is modeled by the process $\left(X_{n}\right)_{0\le n\le N}$
defined by:
\[
\left\{ \begin{array}{lcl}
X_{0} & = & 0\\
X_{n+1} & = & X_{n}+v_{n}S_{n+1}\delta t+L\left(\dfrac{v_{n}}{V_{n+1}}\right)V_{n+1}\delta t,
\end{array}\right.
\]
where $V_{n+1}\delta t$ is the market volume over the period $[t_n,t_{n+1}]$.
We assume that the process $(V_{n})_{n}$ is $\mathcal{F}_{0}$-measurable.

\begin{rem}
In applications, $L$ is often a power function, i.e. $L(\rho)=\eta\left|\rho\right|^{1+\phi}$
with $\phi>0$, or a function of the form $L(\rho)=\eta\left|\rho\right|^{1+\phi}+\psi|\rho|$
with $\phi,\psi>0$. In other words, the execution costs per share are of the form $\eta |\rho|^\phi + \psi$ where $\rho$ is the participation rate. The coefficients for this form of temporary market impact function has been estimated in \cite{almgrenciti} by Almgren et al.
\end{rem}

For the model to be realistic, we also assume that the bank cannot buy/sell too many shares compared to the available liquidity. In other words, we impose that $\underline{\rho} V_{n+1} \le v_n \le \overline{\rho} V_{n+1}$, where $\underline{\rho}$ can be of either sign.\footnote{Constraints of this type are sometimes imposed in the contract.}

We then introduce a non-empty set $\mathcal{N}\subset\lbrace1,\ldots,N-1\rbrace$
corresponding to time indices at which the bank can choose to exercise the option (note that $N\notin\mathcal{N}$
because the bank has no choice at time $T$: the settlement is done at time $T$ if it has not been done beforehand). In practice, this set
is usually of the form $\{n_{0},\ldots,N-1\}$ where $n_{0}>1$.

When the bank decides to exercise the option, or at maturity if the option has not been exercised, the settlement of the contract occurs.
We denote by $n^{\star}\in\mathcal{N}\cup\{N\}$ the time index corresponding to the settlement. At time $t_{n^\star}$, the bank delivers $\frac{F}{A_{n^{\star}}} - Q$ shares to the firm and then the bank has to return to a flat position. Overall, the bank will have to buy $\frac{F}{A_{n^{\star}}}- q_{n^{\star}}$ shares after time $t_{n^\star}$. As this problem is a pure optimal execution problem that only depends on the past through $(q_{n^\star},A_{n^{\star}})$, we can assume that the cost of buying these shares is of the form
$$\left(\frac{F}{A_{n^{\star}}}- q_{n^{\star}}\right)S_{n^{\star}} + \ell\left(\frac{F}{A_{n^{\star}}}- q_{n^{\star}}\right),$$ where the function $\ell$ represents a risk-liquidity premium to account for the execution costs and the risk in the execution process after time $t_{n^\star}$ (see Remark 2 below and \cite{gueant, gueantpov, book}).

The optimization problem the bank faces is given by:

\[
\sup_{(v,n^{\star})\in\mathcal{A}}\mathbb{E}\left[-\exp\left(-\gamma\left(F-X_{n^{\star}}-\left(\frac{F}{A_{n^{\star}}}- q_{n^{\star}}\right)S_{n^{\star}} - \ell\left(\frac{F}{A_{n^{\star}}}- q_{n^{\star}}\right)\right)\right)\right],
\]

where $\mathcal{A}$ is the set of admissible strategies defined as:
\[
\begin{array}{cccl}
\mathcal{A} & = & \Big\{(v,n^{\star})\Big| & v=(v_{n})_{0\le n\le n^{\star}-1}\text{ is }\left(\mathcal{F}
\right)\text{-adapted}, \underline{\rho} V_{n+1} \le v_n \le \overline{\rho} V_{n+1},\\
 &  &  & n^{\star}\text{ is a }\left(\mathcal{F}\right)\text{-stopping time taking values in }\mathcal{N}\cup\{N\}\Big\},
\end{array}
\]
and where $\gamma$ is the absolute risk aversion parameter of the
bank.

\begin{rem}
\label{l}
To be consistent with the expected utility framework, the penalty function $\ell$ should be linked to the indifference price of a block of $q$ shares (which is indeed of the form $qS + \ell(q)$). Expressions for $\ell$ have been given in a continuous time model in \cite{gueant}. If we assume that execution after $t_{n^\star}$ is at constant participation rate, we can also consider for $\ell$ the risk-liquidity premium associated with a liquidation at participation rate $\rho$ (see \cite{gueantpov}):
$$\ell(q) = \frac{L(\rho)}{\rho} |q|  + \frac{\gamma \sigma^2 }{6\rho V} |q|^3,$$ where $V$ is the average value of the market volume process.
This is the function we consider throughout this article -- see also \cite{book}.
\end{rem}

\section{Pricing and optimal strategy}

\subsection{Bellman equations}

Solving this problem requires to determine both the optimal execution
strategy of the bank and the optimal stopping time. For that purpose, we use dynamic programming. We introduce the value functions $(u_n)_{0\le n \le N}$ defined by:

$$u_{n}(x,q,S,A)=$$$$  \sup_{(v,n^{\star})\in\mathcal{A}_{n}}\E\left[-\exp\left(-\gamma\left(F-X_{n^{\star}}^{n,x,v}-\left(\frac{F}{A^{n,A,S}_{n^{\star}}}- q^{n,q,v}_{n^{\star}}\right)S^{n,S}_{n^{\star}} - \ell\left(\frac{F}{A^{n,A,S}_{n^{\star}}}- q^{n,q,v}_{n^{\star}}\right)\right)\right)\right],$$
where $\mathcal{A}_{n}$ is the set of admissible strategies at time
$t_n$ defined as:
$$
\begin{array}{cccl}
\mathcal{A}_{n} & = & \Big\{(v,n^{\star})\Big| & v=(v_{k})_{n\le k\le n^{\star}-1}\text{ is }\left(\mathcal{F}\right)\text{-adapted}, \underline{\rho} V_{k+1} \le v_k \le \overline{\rho} V_{k+1},\\
 &  &  & n^{\star}\text{ is a }\left(\mathcal{F}\right)\text{-stopping time taking values in }\left(\mathcal{N}\cup\{N\}\right)\cap\{n,\ldots,N\}\Big\},
\end{array}
$$
and where the state variables are defined for $0\leq n\leq k\leq N$
and $k>0$ by:
$$
\left\{ \begin{array}{lcl}
X_{k}^{n,x,v} & = & x+\Sum_{j=n}^{k-1}v_{j}S_{j+1}^{n,S}\delta t+L\left(\frac{v_{j}}{V_{j+1}}\right)V_{j+1}\delta t\\
q_{k}^{n,q,v} & = & q+\Sum_{j=n}^{k-1}v_{j}\delta t\\
S_{k}^{n,S} & = & S+\sigma\sqrt{\delta t}\Sum_{j=n}^{k-1}\epsilon_{j+1}\\
A_{k}^{n,A,S} & = & \frac{n}{k}A+\frac{1}{k}\Sum_{j=n}^{k-1}S_{j+1}^{n,S}.
\end{array}\right.
$$

\begin{rem}
In fact, when $n=0$, the function $u_n$ is independent of $A$.
\end{rem}

It is then straightforward to prove that the value functions are characterized by the recursive equations of Proposition \ref{rec}:

\begin{prop}
\label{rec}
The family of functions $(u_{n})_{0\le n\le N}$ defined above is
the unique solution of the Bellman equation:
\begin{eqnarray*}
u_{n}(x,q,S,A) & = & \begin{cases}
-\exp\left(-\gamma\left(F-x-\left(\frac{F}{A}- q\right)S - \ell\left(\frac{F}{A}- q\right)\right)\right) & \text{if }n=N,\\
\begin{array}{rl}
\max\bigg\{ & \tilde{u}_{n,n+1}\left(x,q,S,A\right),\\
 & -\exp\left(-\gamma\left(F-x-\left(\frac{F}{A}- q\right)S - \ell\left(\frac{F}{A}- q\right)\right)\right)\bigg\}
\end{array} & \text{if }n\in\mathcal{N},\\
\tilde{u}_{n,n+1}\left(x,q,S,A\right) & \text{otherwise},
\end{cases}
\end{eqnarray*}
where
\begin{align*}
\tilde{u}_{n,n+1}(x,q,S,A) =  \sup_{\underline{\rho} V_{n+1}\le v \le \overline{\rho} V_{n+1}}\E\left[u_{n+1}\Big(X_{n+1}^{n,x,v},q_{n+1}^{n,q,v},S_{n+1}^{n,S},A_{n+1}^{n,A,S}\Big)\right].
\end{align*}
\end{prop}

Our goal will now be to simplify the problem so that a solution can be computed numerically. By ``solution'', we mean two different things: (i) an optimal strategy for both optimal execution and optimal stopping and (ii) a price for the ASR contract. For the second problem we use the concept of indifference price (see below).

\subsection{Towards a simpler system of equations}

Following the classical economic fact that the current wealth of an agent has no impact on its decision-making process when he has a constant absolute risk aversion, we introduce the change of variables:
\begin{eqnarray*}
u_{n}(x,q,S,A) & = & -\exp\left(-\gamma\left(-x+qS-\theta_{n}\left(q,S,A\right)\right)\right)
\end{eqnarray*}

The functions $(\theta_{n})_n$ are then characterized by the recursive equations of Proposition \ref{theta}:

\begin{prop}
\label{theta}
$\left(\theta_{n}\right)_n$ satisfies:
\begin{eqnarray*}
\theta_{n}(q,S,A) & = & \begin{cases}
F\left(\dfrac{S}{A}-1\right)+\ell\left(\dfrac{F}{A}-q\right) & \text{if }n=N,\\
\min\bigg\{\tilde{\theta}_{n,n+1}\left(q,S,A\right),F\left(\dfrac{S}{A}-1\right)+\ell\left(\dfrac{F}{A}-q\right)\bigg\} & \text{if }n\in\mathcal{N},\\
\tilde{\theta}_{n,n+1}\left(q,S,A\right) & \text{otherwise},
\end{cases}
\end{eqnarray*}
where

\begin{eqnarray*}
\tilde{\theta}_{n,n+1}\left(q,S,A\right) & = & \inf_{\underline{\rho} V_{n+1} \le v \le \overline{\rho} V_{n+1} }\logexpo q\sigma\sqrt{\delta t}\epsilon_{n+1}-L\left(\frac{v}{V_{n+1}}\right)V_{n+1}\delta t\\
 &  & \!\!\!\!\!\!-\theta_{n+1
 }\left(q+v\delta t,S+\sigma\sqrt{\delta t}\epsilon_{n+1},\frac{n}{n+1}A+\frac{1}{n+1}S+\frac{1}{n+1}\sigma\sqrt{\delta t}\epsilon_{n+1}\right)\logexpc.
\end{eqnarray*}
\end{prop}

\begin{rem}
As above for $u_0$, $\theta_0$ is independent of $A$, and we write $\theta_0(q,S)$ instead of $\theta_0(q,S,A)$.
\end{rem}

This change of variables along with the associated recursive equations show that the dimension of the problem can be reduced from 5 to 4. Indeed, the variable $x$ does not appear anymore in the functions $(\theta_n)_n$. In \cite{jaimungal}, and similarly in \cite{asr}, the authors manage, in the case of an ASR contract with fixed number of shares, to reduce the dimensionality of the problem from~5 to~3. They consider indeed respectively the ratio $\frac AS$ and the spread $A-S$ instead of the couple $(A,S)$. In the case of a fixed notional ASR, this dimension reduction is not anymore possible due to the additional convexity in the payoff (the term $\frac{F}{A}$), and the problem to be solved is in dimension 4.

To find the optimal strategy, the method is first to solve recursively the equations for $(\theta_n)_n$. Then, when $n \in \mathcal{N}$, the option is exercised if $\tilde{\theta}_{n,n+1}\left(q,S,A\right)>F\left(\dfrac{S}{A}-1\right)+\ell\left(\dfrac{F}{A}-q\right)$. In the case where the option is not exercised, the optimal strategy consists in sending an order of size $v^* \delta t$, where $v^*$ is a minimizer in the definition of $\tilde{\theta}_{n,n+1}$.

\subsection{Price and strategy: the effects at stake}

In addition to a reduction in the dimension of the problem, the main interest of the change of variables introduced earlier lies in the link between the functions $(\theta_n)_n$ and the indifference price of the ASR contract.

The indifference price of the ASR contract is, by definition, the amount of cash (positive or negative) to give to the bank so that it is indifferent between accepting and not accepting the ASR contract. Mathematically, it is defined by:

$$ \Pi:=\inf \Big\{ p \ \Big| \ \sup_{(v,n^{\star})\in\mathcal{A}}\E\left[-\exp\left(-\gamma\left(p+F-X^{0,0,v}_{n^{\star}}\right.\right.\right.$$$$\left.\left.\left.-\left(\frac{F}{A^{0,S_0,S_0}_{n^{\star}}}- q^{0,0,v}_{n^{\star}}\right)S^{0,S_0}_{n^{\star}} - \ell\left(\frac{F}{A^{0,S_0,S_0}_{n^{\star}}}- q^{0,0,v}_{n^{\star}}\right)\right)\right)\right]\geq -1   \Big\}. $$

The link between the price and the functions $(\theta_n)_n$ is given by the following Proposition:

\begin{prop}
\label{price}
$$\Pi = \theta_0(0,S_0).$$
\end{prop}

When $\Pi$ is positive, it means that the execution costs and the cost of the risk borne by the bank are larger than the potential gain for the bank associated with the option embedded in the contract. On the contrary, if $\Pi$ is negative, the bank values the option sufficiently high to compensate the execution costs and the risk associated with the execution strategy.

\begin{rem}
In practice, when a firm wants to buy back shares, competition occurs between banks willing to enter the deal (banks for which $\Pi \le 0$). However, instead of proposing rebates in cash that would correspond to at most $-\Pi$, banks propose a discount $\beta$ on the average price. The optimization problem is then:
\[
\sup_{(v,n^{\star})\in\mathcal{A}}\mathbb{E}\left[-\exp\left(-\gamma\left(F-X^{0,0,v}_{n^{\star}}-\left(\frac{F}{(1-\beta)A^{0,S_0,S_0}_{n^{\star}}}- q^{0,0,v}_{n^{\star}}\right)S^{0,S_0}_{n^{\star}}\right.\right.\right.$$$$\left.\left.\left. - \ell\left(\frac{F}{(1-\beta)A^{0,S_0,S_0}_{n^{\star}}}- q^{0,0,v}_{n^{\star}}\right)\right)\right)\right].
\]

From a mathematical point of view, this problem is very similar to the original one. However, the maximum rebate $\beta^*$ defined as
\[
\beta^* := \sup\left\lbrace \beta \le 1 | \sup_{(v,n^{\star})\in\mathcal{A}}\mathbb{E}\left[-\exp\left(-\gamma\left(F-X^{0,0,v}_{n^{\star}}-\left(\frac{F}{(1-\beta)A^{0,S_0,S_0}_{n^{\star}}}- q^{0,0,v}_{n^{\star}}\right)S^{0,S_0}_{n^{\star}}\right.\right.\right.\right.\]
\[\left.\left.\left.\left. - \ell\left(\frac{F}{(1-\beta)A^{0,S_0,S_0}_{n^{\star}}}- q^{0,0,v}_{n^{\star}}\right)\right)\right)\right] \ge -1 \right\rbrace,
\]
cannot be computed directly.
\end{rem}

The functions $(\theta_n)_n$ also enable to understand the different effects at stake when it comes to the execution strategy and the choice of the stopping time. For that purpose, let us expand the expression for $\theta_n$ (we omit superscripts for the sake of simplicity). As proved in Appendix~B, we have:
\begin{align*}
\theta_{n}(q,S,A) = &\inf_{(v,n^{\star})\in\mathcal{A}_{n}}\logexpo \underbrace{\sigma\sqrt{\delta t}\sum_{j=n}^{n^{\star}-1}\left(q_{j}-\frac{j}{n^{\star}}\frac{F}{A_{n^{\star}}}\right)\epsilon_{j+1}}_{\text{Risk\ term I}}+\underbrace{\frac{n}{n^{\star}}(A-S)\frac{F}{A_{n^{\star}}}}_{\text{Spread\ term}}\Biggr)\nonumber \\
 & -\underbrace{\sum_{j=n}^{n^{\star}-1}L\left(\frac{v_{j}}{V_{j+1}}\right)V_{j+1}\delta t}_{\text{Liquidity\ term\ I}}-\underbrace{\ell\left(\frac{F}{A_{n^{\star}}} - q_{n^{\star}}\right)}_{\text{ Post-exercise\ risk-liquidity\ term}}\logexpc.
\end{align*}

To understand the different effects, one has to distinguish between what happens before the option is exercised and after the option is exercised.

Before time $t_{n^\star}$, the bank buys back shares on the market. As for all execution problems, there is an execution cost component and a risk component: (i) the bank pays execution costs depending on its participation rate to the market (this is ``Liquidity term I''), and (ii) the bank is exposed to price moves. However, due to the form of the payoff, the risk is partially hedged. When prices go up, the bank buys shares at higher prices but the number of shares to buy ($\frac{F}{A}$) decreases. Similarly the other way round, when prices go down, the bank buys shares at lower prices but the number of shares to buy increases. Contrary to what happens with ASR contracts with fixed number of shares, the hedge is only partial in the case of a fixed notional ASR: ``Risk term I'' cannot be set to 0 using an adapted strategy.

After time $t_{n^\star}$, the execution costs are similar but the risk is not anymore partially hedged: this is the ``Post-exercise\ risk-liquidity\ term'' (it requires to choose $\ell$ properly, see Remark 2).

The above analysis is linked to the execution process used for buying back shares. Coming to the option, the bank has an incentive to exercise it when $S$ is below $A$, since $\frac{F}{A}$ will mechanically go up, as $A$ goes towards $S$. The cost of not exercising now is somehow given by the ``Spread term''.

We see therefore that there is a trade-off between exercising when $S$ goes below $A$ in order to eventually deliver less shares, and not executing too soon in order to (partially) hedge the risk associated with the execution process. This implies accelerating the buying process when $S$ decreases in order to have less to hedge, and decelerating the buying process (or even selling shares) when $S$ increases.

\section{Numerical methods and examples}

\subsection{A tree-based method}
To solve the problem numerically, we introduce a tree-based method.\footnote{This method is different from the one used in \cite{asr} where nodes were indexed by the spread $S-A$.} This method uses a tree to model the diffusion of prices. For each node in the tree, indexed by a time index $n$ and a price $S$, we shall compute the values of the function $\theta_n(\cdot,S,\cdot)$ on a grid.

\subsubsection*{Structure of the tree}

We consider that innovations $\left(\epsilon_{n}\right)_{n\ge1}$
have the following distribution:%
\footnote{The distribution is chosen to match the first four moments of a standard
normal:
\[
\left\{ \begin{array}{ccc}
\mathbb{E}\left[\epsilon_{n}\right] & = & 0\\
\mathbb{E}\left[\epsilon_{n}^{2}\right] & = & 1\\
\mathbb{E}\left[\epsilon_{n}^{3}\right] & = & 0\\
\mathbb{E}\left[\epsilon_{n}^{4}\right] & = & 3
\end{array}\right.
\]
}
\begin{eqnarray*}
\epsilon_{n} & = & \begin{cases}
+2 & \text{with probability }\frac{1}{12},\\
+1 & \text{with probability }\frac{1}{6},\\
0 & \text{with probability }\frac{1}{2},\\
-1 & \text{with probability }\frac{1}{6},\\
-2 & \text{with probability }\frac{1}{12}.
\end{cases}
\end{eqnarray*}

We introduce, for $n \in \lbrace 0, \ldots, N \rbrace$, and for $\zeta \in \lbrace 0, \ldots, 4n \rbrace$ the function $\Theta_{n}^{\zeta}$ defined by:
\begin{eqnarray*}
\Theta_{n}^{\zeta}(q,A) & = & \theta_{n}\left(q,S_{0}+\sigma\sqrt{\delta t}\left(\zeta-2n\right),A\right).
\end{eqnarray*}

Following Proposition \ref{theta}, the family of functions $\left(\Theta_{n}^{\zeta}\right)$ satisfies:
\begin{eqnarray*}
\Theta_{n}^{\zeta}(q,A) & = & \begin{cases}
F\left(\dfrac{S_{0}+\sigma\sqrt{\delta t}\left(\zeta-2n\right)}{A}-1\right)+\ell\left(\dfrac{F}{A}-q\right) & \text{if }n=N,\\
\min\bigg\{\tilde{\Theta}_{n,n+1}^{\zeta}(q,A),F\left(\dfrac{S_{0}+\sigma\sqrt{\delta t}\left(\zeta-2n\right)}{A}-1\right)+\ell\left(\dfrac{F}{A}-q\right)\bigg\} & \text{if }n\in\mathcal{N},\\
\tilde{\Theta}_{n,n+1}^{\zeta}(q,A) & \text{otherwise},
\end{cases}
\end{eqnarray*}
where
\begin{eqnarray*}
\tilde{\Theta}_{n,n+1}^{\zeta}(q,A) & = & \inf_{v\in\left[\underline{\rho}V_{n+1},\overline{\rho}V_{n+1}\right]}\logexpo q\sigma\sqrt{\delta t}\epsilon_{n+1}-L\left(\frac{v}{V_{n+1}}\right)V_{n+1}\delta t\\
 &  & -\Theta_{n}^{\zeta+\left(\epsilon_{n+1}+2\right)}\biggl(q+v\delta t,\frac{n}{n+1}A\\
 &  & \quad+\frac{1}{n+1}\left(S_{0}+\sigma\sqrt{\delta t}\left(\zeta-2n\right)\right)+\frac{1}{n+1}\sigma\sqrt{\delta t}\epsilon_{n+1}\biggr)\logexpc.
\end{eqnarray*}

The functions $\left(\Theta_{n}^{\zeta}\right)_{n,\zeta}$ are computed recursively on a grid. Each index $(n,\zeta)$ corresponds to a node of the pentanomial tree.

\subsubsection*{Structure of the $(q,A)$-grid}

At each node of the tree, we compute the values of $\Theta_{n}^{\zeta}(q,A)$ for $(q,A) \in \mathcal{G}_{q}\times \mathcal{G}_{A}$, where $\mathcal{G}_{q}$ is a grid of the form $\left\{ \frac{k}{n_q-1} q_{\text{max}} | k \in \lbrace 0, \ldots, n_q-1 \rbrace\right\} $ and where $\mathcal{G}_{A}$ is a grid of the form  $\left\{S_{0} + \xi\left(\frac k{n_A-1} - \frac 12\right)  \sigma\sqrt{N\delta t}| k \in \lbrace 0,\ldots ,n_A-1 \rbrace\right\}$.

For each grid, we need to choose the number of points in the grid ($n_q$ and $n_A$) and the width of the grid (linked to $q_{\text{max}}$ and $\xi$).

The number of shares to deliver is linked to the ratio $\frac{F}{A}$. Consequently, it is not bounded and we need to choose a minimal value for $A$. Since $\mathbb{V}\left[A_{N}\right]\simeq\dfrac{1}{3}N\sigma^{2}\delta t$, it is natural to consider values of $A$ in a range of the form $\left[S_0- \frac \omega {\sqrt 3} \sigma\sqrt{N\delta t}, S_0+ \frac \omega {\sqrt 3} \sigma\sqrt{N\delta t}\right]$ where $\omega$ is the number of standard deviations we want to keep. In practice, we choose $\xi = \frac 2 {\sqrt 3}\omega = 3$, that is around~$2.6$ standard deviations. Then, it is natural for $q_{\text{max}}$, in coherence with the lower bound for~$A$, to consider a value $q_{\text{max}} \simeq \frac{F}{S_0 - \frac \xi 2  \sigma\sqrt{N\delta t}}$.\\

In our numerical method, we consider the optimization of the execution strategy on the grid, that is:
$$
\tilde{\Theta}_{n,n+1}^{\zeta}(q,A) =  \min_{q_{\text{target}}\in\left[q+\underline{\rho}V_{n+1}\delta t,q+\overline{\rho}V_{n+1}\delta t\right]\cap\mathcal{G}_{q}}$$
$$\logexpo q\sigma\sqrt{\delta t}\epsilon_{n+1}- L\left(\frac{q_{\text{target}}-q}{V_{n+1}\delta t}\right)V_{n+1}\delta t  -\Theta_{n}^{\zeta+\left(\epsilon_{n+1}+2\right)}\biggl(q_{\text{target}},\frac{n}{n+1}A$$$$\quad+\frac{1}{n+1}\left(S_{0}+\sigma\sqrt{\delta t}\left(\zeta-2n\right)\right)+\frac{1}{n+1}\sigma\sqrt{\delta t}\epsilon_{n+1}\biggr)\logexpc.$$

\begin{rem}
In practice, if $(V_n)_n$ is a constant process, it is better to have $\underline{\rho}V\delta t$ and
$\overline{\rho}V\delta t$ on the $q$-grid.
\end{rem}

For computing the minimum in the previous equation, and hence for computing $\tilde{\Theta}_{n,n+1}^{\zeta}(q,A)$, we need the values
of $\Theta_{n}^{\zeta+\left(\epsilon_{n+1}+2\right)}\biggl(q_{\text{target}},A'\biggr)$
for $$A'=\frac{n}{n+1}A+\frac{1}{n+1}\left(S_{0}+\sigma\sqrt{\delta t}\left(\zeta-2n\right)\right)+\frac{1}{n+1}\sigma\sqrt{\delta t}\epsilon_{n+1}.$$ As $A'$ does not necessary lie on the grid $\mathcal{G}_{A}$, we use interpolation with natural cubic splines and, if necessary, we extrapolate linearly the functions outside of the domain (for $A$).\\

\subsubsection*{Determination of the optimal strategy}

Using the above recursive equations, we can approximate the values of the functions $(\theta_n)_n$. Moreover, at each step in time and for each combination of the tuple $(q,S,A)$ for which we computed the value, we can easily compute the optimal strategy. We obtain $q^*_{\text{target}}$ from the minimization procedure that permits to compute $\tilde{\Theta}_{n,n+1}^{\zeta}(q,A)$. If $$\tilde{\Theta}_{n,n+1}^{\zeta}(q,A) > F\left(\dfrac{S_{0}+\sigma\sqrt{\delta t}\left(\zeta-2n\right)}{A}-1\right)+\ell\left(\dfrac{F}{A}-q\right),$$ and if we are allowed to exercise, then we exercise the option. Otherwise, we send an order of size $q^*_{\text{target}}-q$ to the market.

\subsection{Examples}

\subsubsection{Reference scenarios}

We now turn to the practical use of our numerical method on examples. We consider the following reference case, that corresponds to rounded values for a major stock of the CAC 40 Index. This case will be used throughout the remainder of this text.

\subsubsection*{Parameters for the stock}
\begin{itemize}
\item $S_{0}=45$ €.
\item $\sigma=0.6$ €$\cdot\text{day}^{-1/2}$, which corresponds to an annual volatility approximately equal to $21\%$.
\item $V=4\ 000\ 000$ stocks$\cdot$ $\text{day}^{-1}$.
\item $L(\rho)=\eta|\rho|^{1+\phi}$ with $\eta=0.1$ € $\cdot\mbox{stock}^{-1}\cdot\text{day}^{-1}$
and $\phi=0.75$.
\end{itemize}

\subsubsection*{Parameters for the ASR contract}
\begin{itemize}
\item $T=63$ days.
\item The set of possible dates for early exercising the option is $\mathcal{N} = [22,62]\cap \mathbb N$.
\item $F=900\ 000\ 000$€.
\end{itemize}

\subsubsection*{Parameters for the bank}
\begin{itemize}
\item Participation rates are bounded by $\underline{\rho} = -25\%$ and $\overline{\rho} = 25\%$ .
\item Risk aversion is $\gamma=2.5\times 10^{-7}$ €$^{-1}$.
\item $\ell$ is as described in Remark 2. In other words, after the option has been exercised, we assume that execution occurs at constant participation rate of $25\%$.
\end{itemize}

\subsubsection*{Parameters for the numerical method}
\begin{itemize}
\item $q_{\text{max}} = 25\ 000\ 000$ stocks.
\item $n_q = 201$.
\item $\xi=3$, corresponding to $\omega \simeq 2.6$ standard deviations.
\item $n_A = 21$.
\end{itemize}

We first start with three examples corresponding to three trajectories for the price.

\begin{figure}[H]
\centering{}\includegraphics[width=0.75\textwidth]{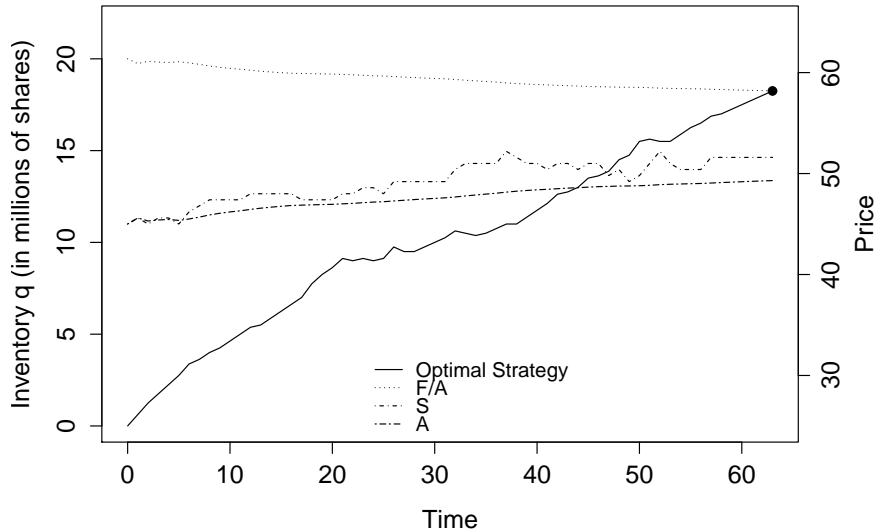}\caption{Optimal strategy when prices are mainly going up.}
\label{ref_up}
\end{figure}

The first example, on Figure \ref{ref_up}, corresponds to an increasing trend and the stock price is therefore above its average. As explained in the previous section, there is no reason in that case to early exercise the option as $A$ increases (the dot corresponds to the date when the option is exercised).

\begin{figure}[H]
\centering{}\includegraphics[width=0.75\textwidth]{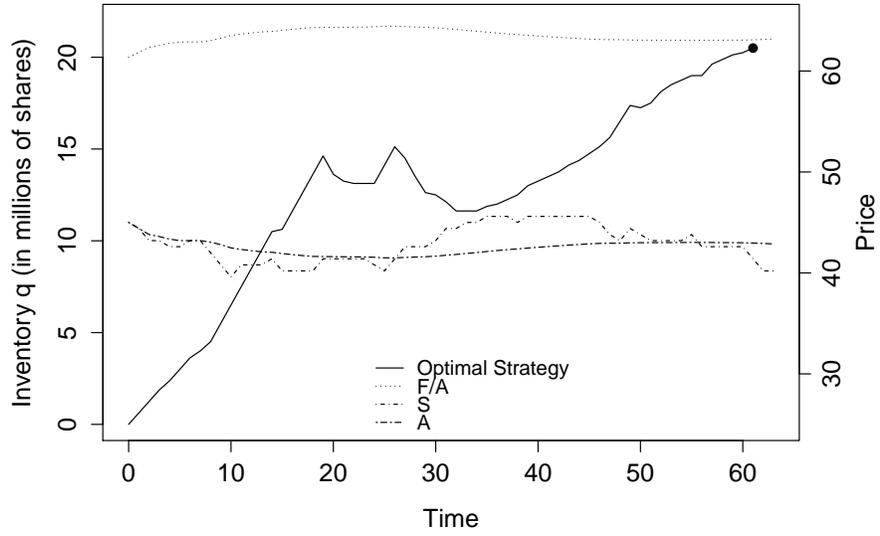}\caption{Optimal strategy when prices are mainly oscillating.}
\label{ref_mid}
\end{figure}

The second example, on Figure \ref{ref_mid}, corresponds to a stock price oscillating around its average value. We see, as expected, that the buying process accelerates when the stock price decreases and decelerates when the stock price increases. We even see that the buying process can turn into a selling process as prices increase. The rationale for that is hedging, and it can be seen on the term we called ``Risk term I'': $\sum_{j=n}^{n^{\star}-1}\left(q_{j}-\frac{j}{n^{\star}}\frac{F}{A_{n^{\star}}}\right)\epsilon_{j+1}$. As prices increase, we expect the number of shares to be eventually bought (that is $\frac{F}{A_{n^{\star}}}$) to decrease and the exercise date (\emph{i.e.} $t_{n^\star}$) to be later than initially thought. Therefore, the term $\frac{j}{n^{\star}}\frac{F}{A_{n^{\star}}}$ decreases and $q$ has to go down in order to reach the same level of risk.
We also see on this second example that the option is exercised near maturity.

\begin{figure}[H]
\centering{}\includegraphics[width=0.75\textwidth]{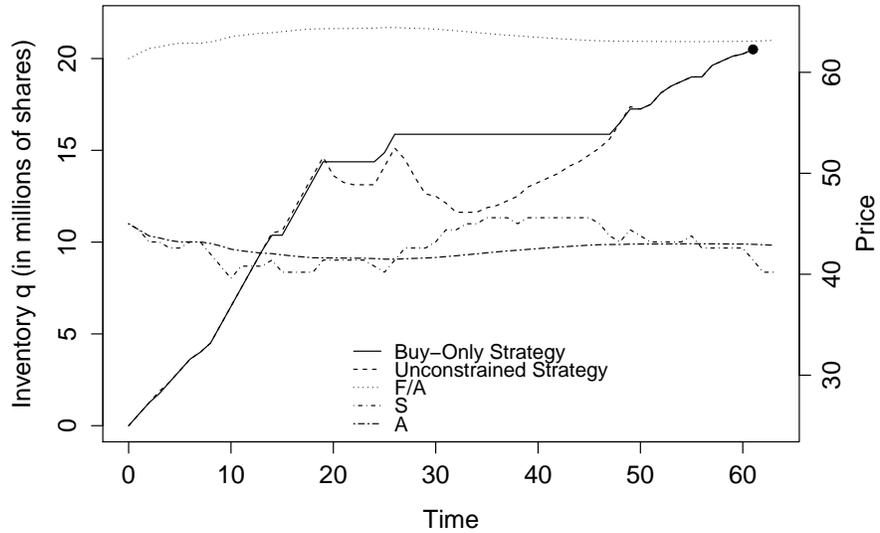}\caption{Optimal strategy when prices are mainly oscillating ($\underline{\rho}=0$).}
\label{constr_mid}
\end{figure}

As the optimal strategy leads to buying and selling shares, it is interesting to understand what happens when we force the bank to use a buy-only strategy. If we impose $\underline{\rho}=0$, we obtain the optimal strategy of Figure \ref{constr_mid} that is substantially different.

\begin{figure}[H]
\centering{}\includegraphics[width=0.75\textwidth]{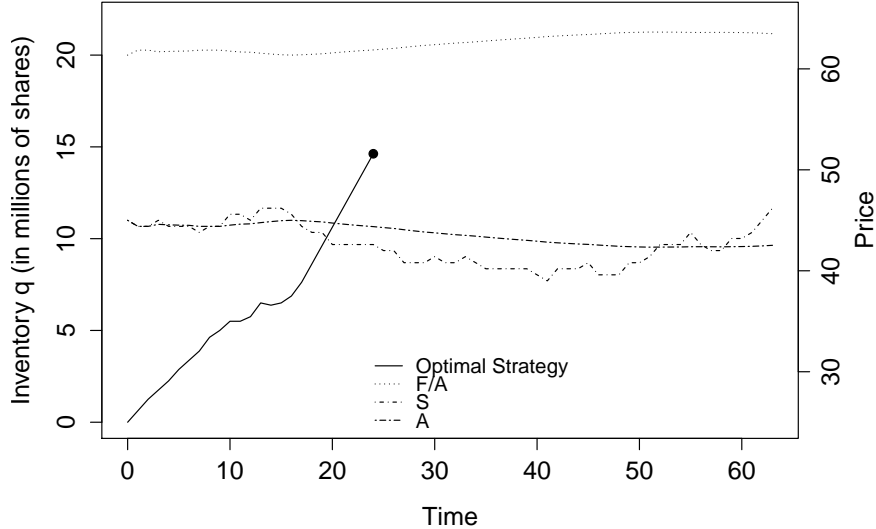}\caption{Optimal strategy when prices are mainly going down.}
\label{ref_down}
\end{figure}

The third trajectory we consider corresponds to decreasing prices (see Figure \ref{ref_down}). In that case, the bank buys shares quite rapidly and exercises  the option as soon as it can, although it has not yet bought the required number of shares. By exercising the option, the bank wants in fact to avoid the natural decrease in $A$ that would lead to an increase in the number of shares it would have to buy.

In addition to optimal strategies, we can compute in our reference case, the price $\Pi$ of the fixed notional ASR contract. Here, this price is negative as we found $\Pi =  -10669023 \simeq - 1.185\% F $. This means, in utility terms, that the gain associated with the optionality component of the ASR contract is important enough to compensate (in utility terms) the execution costs and the risk of the contract. In the case where we impose the use of a buy-only strategy, we obtain $\Pi =  -10330135 \simeq - 1.148\% F$.

\subsubsection{Comparative statics}

In order to understand the role of the main parameters, we carry out comparative statics. We focus on the liquidity of the stock, through the parameter $\eta$, on the volatility of the stock price, through $\sigma$, and on the risk aversion parameter $\gamma$.

\subsubsection*{Effect of execution costs}

\begin{figure}[H]
\centering{}\includegraphics[width=0.75\textwidth]{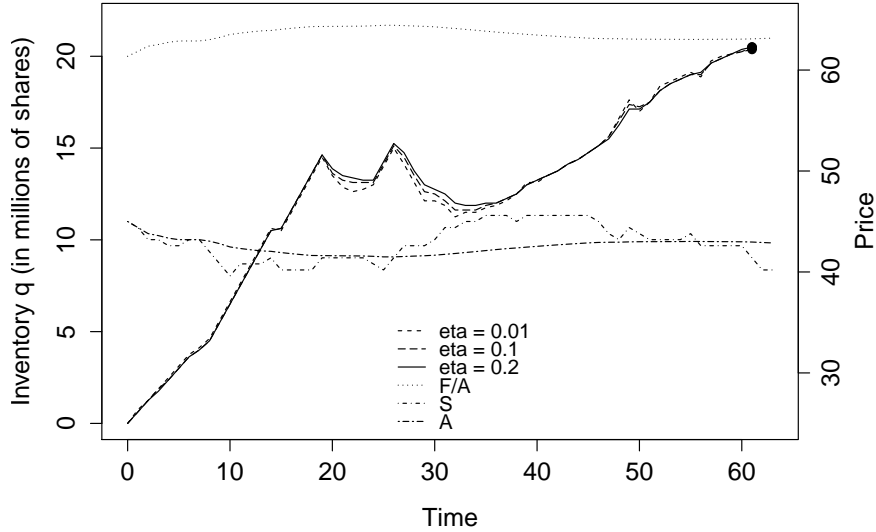}\caption{Optimal strategy when prices are mainly oscillating, for different values of $\eta$.}
\label{eta_mid}
\end{figure}

Regarding execution costs, we consider our reference case with 3 values for the parameter $\eta$: $0.01, 0.1,$ and $0.2$. We concentrate on the second price trajectory (Figure \ref{eta_mid}) as it shows very well the role of $\eta$: the more liquid the stock, the larger the number of round trips on the stock for hedging purposes.

The differences in terms of prices are the following:

\begin{table}[H]
\begin{centering}
\begin{tabular}{|c|c|c|c|}
\hline
$\eta$ & $0.01$ & $0.1$ & $0.2$\tabularnewline
\hline
$\frac{\Pi}{F}$ & $ -1.254\% $  & $-1.185\%  $ & $-1.117\% $\tabularnewline
\hline
\end{tabular}
\par\end{centering}
\end{table}

As expected, the more liquid the stock, the lower the indifference price: the buying process is indeed less costly in itself, and round trips on the stock to hedge risk are also less costly.

\subsubsection*{Effect of volatility}

Let us come to the effect of volatility. We consider our reference case with 3 values for the parameter $\sigma$: $0.3, 0.6,$ and $1.2$. Two effects are at stake. On the one hand, the more volatile the stock, the more important the magnitude of the terms we called ``Risk term I'' and ``Post-exercise\ risk-liquidity\ term''. On the other hand, the larger $\sigma$, the more valuable the option for the bank.

In terms of prices, we see that the first effect dominates as $\Pi$ is an increasing function of $\sigma$. We also see that $\sigma$ is a very important driver of the price of an ASR contract.

\begin{table}[H]
\begin{centering}
\begin{tabular}{|c|c|c|c|}
\hline
$\sigma$ & $0.3$ & $0.6$ & $1.2$\tabularnewline
\hline
$\frac{\Pi}{F}$ & $ -2.163\% $  & $-1.185\% $ & $-0.605\%$\tabularnewline
\hline
\end{tabular}
\par\end{centering}

\end{table}

\subsubsection*{Effect of risk aversion}

\begin{figure}[H]
\centering{}\includegraphics[width=0.75\textwidth]{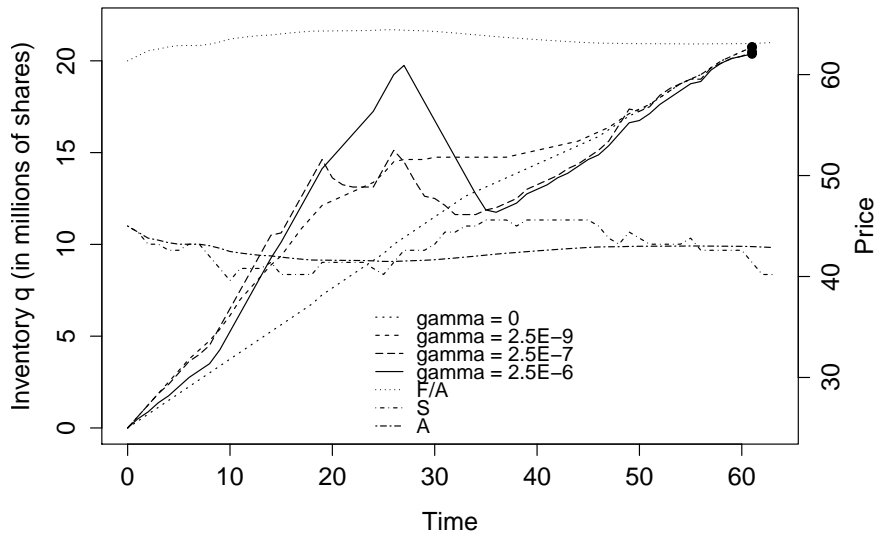}\caption{Optimal strategy when prices are mainly oscillating, for different values of $\gamma$.}
\label{gamma_mid}
\end{figure}

We go on with risk aversion. We consider our reference case with 4 values for the parameter $\gamma$:\footnote{We can extend the results to the case $\gamma=0$.} $0, 2.5\times10^{-9}, 2.5\times10^{-7},$ and $2.5\times10^{-6}$. Figure \ref{gamma_mid} shows the complex influence of $\gamma$ on the optimal strategy for the second stock price trajectory where prices mainly oscillate.

The first thing we see is that, when $\gamma = 0$, the bank buys almost the same number of shares everyday, the changes being due to changes in the targeted number of shares to be bought (as $A$ oscillates). It is also important to understand that the reason why the bank buys shares instead of waiting is because we consider execution at a participation rate of $25\%$ after the exercise date.

In the case of a risk-averse agent, we see different behaviors depending on the level of risk aversion, because there are several effects at stake. The randomness is indeed in both the execution process and the number of shares to buy. On the one hand, risk aversion encourages to exercise the option as soon as $S$ is below $A$ to cut the latter type of risk. However, exercising early reduces the period during which the former risk is partially hedged. The type of risk that is more important to hedge depends on the value of~$\gamma$. This is why we observe so much variety in the optimal strategy. Choosing the value of~$\gamma$ corresponding to the risk the bank wants to bear is therefore very important.\footnote{This is a classical problem in models involving an expected utility. It is encountered very often in optimal execution (for instance in models à la Almgren-Chriss) and in asset management (for instance in the Markowitz approach, or in models à la Merton).}

\newpage

What is however very clear is the influence of $\gamma$ on the price of the ASR contract: the more risk averse the bank, the less it values entering into an ASR contract with a firm.

\begin{table}[H]
\begin{centering}
\begin{tabular}{|c|c|c|c|c|}
\hline
$\gamma$ & $0$ & $2.5\times10^{-9}$ & $2.5\times10^{-7}$ & $2.5\times10^{-6}$\tabularnewline
\hline
$\frac{\Pi}{F}$ & $ -1.499\%  $ & $-1.490\%   $ & $-1.185\%  $ & $-0.468\% $\tabularnewline
\hline
\end{tabular}
\par\end{centering}
\end{table}

\section*{Conclusion}

This paper is a contribution to a new literature on pricing inspired from the literature on optimal execution. Specifically, we have presented in this article a model for characterizing and computing the optimal strategy of a bank entering a fixed notional ASR contract with a firm. Our discrete-time model enables to model the interactions between the execution problem linked to an ASR and the Asian/Bermudean option that is part of the contract, whereas classical approaches would separate the two questions. In addition to providing an optimal strategy, we define an indifference price for the contract. The numerical method we developed for approximating solutions works very well in practice and provides both optimal strategies and prices.

\section*{Appendix A: Introduction of permanent market impact}

In the setup presented above, market impact was only temporary
and boiled down to execution costs.\footnote{We believe indeed that there is little permanent impact of trades in the case of an ASR contract, because buy-back contracts are very often announced in advance.} Here, we briefly show how to generalize
the previous model in order to introduce a linear form of permanent market impact (see \cite{gatheral} for the rationale underlying a linear form of market impact).

In this appendix, we assume that $S_{n+1}$ is the
closing price of the $n^{th}$ day and that the average price in the payoff of the ASR contract is the average of the closing prices.

The dynamics of the portfolio is the same:
\[
\left\{ \begin{array}{lcl}
q_{0} & = & 0\\
q_{n+1} & = & q_{n}+v_{n}\delta t.
\end{array}\right.
\]

The dynamics of the price is however impacted by the strategy of the bank:
$$
\forall n\in\lbrace0,\ldots,N-1\rbrace,\quad S_{n+1}=S_{n}+\sigma\sqrt{\delta t}\epsilon_{n+1} + kv_n \delta t,
$$
where the additional term means that the price goes up when the bank buys shares and goes down when the bank sells shares, as in the classical Almgren-Chriss framework.

The average price is still
$$
A_{n} =\frac{1}{n}\sum_{k=1}^{n}S_{k}.
$$

If we assume that the order of size $v_{n}\delta t$ is executed evenly over the interval $[t_n,t_{n+1}]$ (in a continuous time Almgren-Chriss model), this
leads to the following dynamics for the cash account:
$$
X_{n+1}=X_{n}+S_{n+1}v_{n}\delta t - L\left(\frac{v_{n}}{V_{n+1}}\right)V_{n+1}\delta t - \frac{k}{2}(q_{n+1}-q_n)^2-\frac{\sigma v_{n}\delta t^{\frac{3}{2}}}{\sqrt{3}}\epsilon'_{n+1},
$$
where $\left((\epsilon_{k},\epsilon'_{k}\right))_{k}$ are i.i.d.
random variables with moment generating function defined on $\R_+$, with $\mathbb{E}\left[(\epsilon_{k},\epsilon'_{k}\right)]=0$ and
$\mathbb{V}[(\epsilon_{k},\epsilon'_{k})]=\left(\begin{array}{cc}
1 & \dfrac{\sqrt{3}}{2}\\
\dfrac{\sqrt{3}}{2} & 1
\end{array}\right)$.

\begin{rem}
The term $\frac{k}{2}(q_{n+1}-q_n)^2$ comes from the fact that the expected value of the price paid for the shares is the average between $S_n$ and $S_n + k(q_{n+1}-q_n)$. The noise term $\frac{\sigma v_{n}\delta t^{\frac{3}{2}}}{\sqrt{3}}\epsilon'_{n+1}$
corresponds to the fact that we execute progressively and evenly over the day while the price
$S_{n+1}$ is the closing price.
\end{rem}

Permanent market impact also has to be taken into account in the price paid for the shares after time $t_{n^\star}$. For that purpose, we replace $\ell(q)$ by $\ell(q) + \frac k2 q^2$. Hence, the optimization problem the bank faces becomes:
$$
\sup_{(v,n^{\star})\in\mathcal{A}}\E\left[-\exp\left(F-X_{n^{\star}}-\left(\frac{F}{A_{n^{\star}}}- q_{n^{\star}}\right)S_{n^{\star}} - \frac k2 \left(\frac{F}{A_{n^{\star}}}- q_{n^{\star}}\right)^2 - \ell\left(\frac{F}{A_{n^{\star}}}- q_{n^{\star}}\right)\right)\right].
$$

The problem can be solved recursively using the Bellman equations:
$$
u_{n}(x,q,S,A)=
$$$$\begin{cases}
-\exp\left(-\gamma\left(F-x-\left(\frac{F}{A}- q\right)S - \frac k2 \left(\frac{F}{A}- q\right)^2 - \ell\left(\frac{F}{A}- q\right)\right)\right) & \text{if }n=N,\\
\max\bigg\{\tilde{u}_{n,n+1}\left(x,q,S,A\right),\\
-\exp\left(-\gamma\left(F-x-\left(\frac{F}{A}- q\right)S - \frac k2 \left(\frac{F}{A}- q\right)^2 - \ell\left(\frac{F}{A}- q\right)\right)\right)\bigg\}, & \text{if }n\in\mathcal{N},\\
\tilde{u}_{n,n+1}\left(x,q,S,A\right) & \text{otherwise},
\end{cases}
$$
where
\begin{eqnarray*}
\tilde{u}_{n,n+1}(x,q,S,A) & = & \sup_{v\in\R}\E\Big[u_{n+1}\Big(X_{n+1},q_{n+1},S_{n+1},A_{n+1}\Big)\Big].\label{eq:def-utilde2}
\end{eqnarray*}
If we define
\begin{eqnarray*}
u_{n}(x,q,S,A) & = & -\exp\left(-\gamma\left(-x+qS-\theta_{n}\left(q,S,A\right)\right)\right),
\end{eqnarray*}
the recursive equations for $\theta_{n}$ given in Proposition \ref{theta} become:
\begin{eqnarray*}
\theta_{n}(q,S,A) & = & \begin{cases}
F\left(\dfrac{S}{A}-1\right)+\frac k2 \left(\dfrac{F}{A}-q\right)^2 + \ell\left(\dfrac{F}{A}-q\right) & \text{if }n=N,\\
\min\bigg\{\tilde{\theta}_{n,n+1}\left(q,S,A\right),F\left(\dfrac{S}{A}-1\right)+\frac k2\left(\dfrac{F}{A}-q\right)^2+ \ell\left(\dfrac{F}{A}-q\right)\bigg\} & \text{if }n\in\mathcal{N},\\
\tilde{\theta}_{n,n+1}\left(q,S,A\right) & \text{otherwise},
\end{cases}
\end{eqnarray*}
where
\begin{eqnarray*}
&&\tilde{\theta}_{n,n+1}\left(q,S,A\right)\\
& = & \inf_{\underline{\rho} V_{n+1} \le v \le \overline{\rho} V_{n+1} }\logexpo q\sigma\sqrt{\delta t}\epsilon_{n+1} + kvq\delta t + \frac{\sigma v\delta t^{\frac{3}{2}}}{\sqrt{3}}\epsilon'_{n+1}\\
&& + \frac{k}2 v^2 \delta t^2 -L\left(\frac{v}{V_{n+1}}\right)V_{n+1}\delta t -\theta_{n+1
 }\Bigg(q+v\delta t,S+kv\delta t+ \sigma\sqrt{\delta t}\epsilon_{n+1},\\
 &&\frac{n}{n+1}A+\frac{1}{n+1}S+ \frac{1}{n+1} k v\delta t+\frac{1}{n+1}\sigma\sqrt{\delta t}\epsilon_{n+1}\Bigg)\logexpc.
\end{eqnarray*}

In that case, approximations with splines need to be used for $S$ and $A$. The problem is therefore slightly more complicated from a numerical point of view, but our framework can be used even in the case when one wants to model permanent market impact.

\section*{Appendix B: Proofs}

\footnotesize{

\textbf{Proof of Proposition \ref{theta}:}

The case $n = N$ is straightforward.

When $n<N$ and $n\notin \mathcal N$, we have:
\begin{eqnarray*}
&&\theta_{n}(q,S,A)\\
&= & \dfrac{1}{\gamma}\log\left(-\sup_{\underline{\rho} V_{n+1}\le v \le \overline{\rho} V_{n+1}}\mathbb{E}\left(u_{n+1}\Big(X_{n+1}^{n,x,v},q_{n+1}^{n,q,v},S_{n+1}^{n,S},A_{n+1}^{n,A,S}\Big)\right)\right)-x+qS\\
 & = & \inf_{\underline{\rho} V_{n+1}\le v \le \overline{\rho} V_{n+1}}\logexpo -X_{n+1}^{n,x,v}+x+q_{n+1}^{n,q,v}S_{n+1}^{n,S}-qS-\theta_{n+1}\left(q_{n+1}^{n,q,v},S_{n+1}^{n,S},A_{n+1}^{n,A,S}\right)\logexpc\\
 & = & \inf_{\underline{\rho} V_{n+1}\le v \le \overline{\rho} V_{n+1}}\logexpo -v(S + \sigma \epsilon_{n+1} \sqrt{\delta t})\delta t - L\left(\frac{v}{V_{n+1}}\right)V_{n+1}\delta t \\
 &&+(q+v\delta t)(S + \sigma \epsilon_{n+1} \sqrt{\delta t})-qS-\theta_{n+1}\Big(q+v\delta t,S+\sigma\sqrt{\delta t}\epsilon_{n+1},\\
 &&\frac{n}{n+1}A+\frac{1}{n+1}S+\frac{1}{n+1}\sigma\sqrt{\delta t}\epsilon_{n+1}\Big)\logexpc\\
&=& \tilde{\theta}_{n,n+1}\left(q,S,A\right).
\end{eqnarray*}

For the last case, we proceed in the same way.\qed

\textbf{Proof of Proposition \ref{price}:}

By definition, we have $ -\exp(\gamma \Pi) = u_0(0,0,S_0) = -\exp(\gamma \theta_0(0,S_0))$. Hence the result.\qed

\textbf{Proof of the expression for $\theta_n$:}

We have:
\begin{eqnarray*}
&&u_{n}(x,q,S,A)\\
&=& -\exp\left(-\gamma\left(-x+qS-\theta_{n}\left(q,S,A\right)\right)\right)\\ &=&\sup_{(v,n^{\star})\in\mathcal{A}_{n}}\E\left[-\exp\left(-\gamma\left(-X_{n^{\star}}^{n,x,v}-\left(\frac{F}{A^{n,A,S}_{n^{\star}}}- q^{n,q,v}_{n^{\star}}\right)S^{n,S}_{n^{\star}} - \ell\left(\frac{F}{A^{n,A,S}_{n^{\star}}}- q^{n,q,v}_{n^{\star}}\right)\right)\right)\right]\\
&=&\sup_{(v,n^{\star})\in\mathcal{A}_{n}}\E\left[-\exp\left(-\gamma\left(-X_{n^{\star}}^{n,x,v}+ q^{n,q,v}_{n^{\star}}S^{n,S}_{n^{\star}} +\frac{F}{A^{n,A,S}_{n^{\star}}}\left(A^{n,A,S}_{n^{\star}} - S^{n,S}_{n^{\star}}\right) - \ell\left(\frac{F}{A^{n,A,S}_{n^{\star}}}- q^{n,q,v}_{n^{\star}}\right)\right)\right)\right].\\
\end{eqnarray*}

Now, by definition:

\begin{eqnarray*}
&&-X_{n^{\star}}^{n,x,v}+ q^{n,q,v}_{n^{\star}}S^{n,S}_{n^{\star}}\\
&=&-x - \sum_{j=n}^{n^{\star}-1} v_j S^{n,S}_{j+1}\delta t - \sum_{j=n}^{n^{\star}-1}L\left(\frac{v_{j}}{V_{j+1}}\right)V_{j+1}\delta t + q^{n,q,v}_{n^{\star}}S^{n,S}_{n^{\star}}\\
&=& -x + qS + \sigma \sqrt{\delta t} \sum_{j=n}^{n^{\star}-1} q^{n,q,v}_j \epsilon_{j+1} - \sum_{j=n}^{n^{\star}-1}L\left(\frac{v_{j}}{V_{j+1}}\right)V_{j+1}\delta t.\\
\end{eqnarray*}

Consequently:

\begin{align*}
\theta_{n}(q,S,A) = &\inf_{(v,n^{\star})\in\mathcal{A}_{n}}\logexpo \sigma\sqrt{\delta t}\sum_{j=n}^{n^{\star}-1}q^{n,q,v}_{j}\epsilon_{j+1}+ \frac{F}{A^{n,A,S}_{n^{\star}}}\left(A^{n,A,S}_{n^{\star}} - S^{n,S}_{n^{\star}}\right)\\
 & -\sum_{j=n}^{n^{\star}-1}L\left(\frac{v_{j}}{V_{j+1}}\right)V_{j+1}\delta t - \ell\left(\frac{F}{A^{n,A,S}_{n^{\star}}} - q^{n,q,v}_{n^{\star}}\right)\logexpc.
\end{align*}

Now, straightforward computations lead to:

$$A^{n,A,S}_{n^{\star}} - S^{n,S}_{n^{\star}} = \frac{n}{n^{\star}}(A-S)  - \sigma\sqrt{\delta t}\sum_{j=n}^{n^{\star}-1}\frac{j}{n^{\star}}\epsilon_{j+1}.$$

Hence, we get:

\begin{align*}
\theta_{n}(q,S,A) = &\inf_{(v,n^{\star})\in\mathcal{A}_{n}}\logexpo \sigma\sqrt{\delta t}\sum_{j=n}^{n^{\star}-1}\left(q^{n,q,v}_{j}-\frac{j}{n^{\star}}\frac{F}{A^{n,A,S}_{n^{\star}}}\right)\epsilon_{j+1}+\frac{n}{n^{\star}}(A-S)\frac{F}{A^{n,A,S}_{n^{\star}}}\Biggr)\nonumber \\
 & -\sum_{j=n}^{n^{\star}-1}L\left(\frac{v_{j}}{V_{j+1}}\right)V_{j+1}\delta t-\ell\left(\frac{F}{A^{n,A,S}_{n^{\star}}} - q^{n,q,v}_{n^{\star}}\right)\logexpc.
\end{align*}\qed

}

\normalsize{

\bibliographystyle{plain}

\begin{thebibliography}{10}

\bibitem{almgren}
R.~Almgren and N.~Chriss.
\newblock Value under liquidation.
\newblock {\em Risk}, 12(12):61--63, 1999.

\bibitem{almgren2}
R. Almgren, N. Chriss, Optimal execution of portfolio transactions. Journal of Risk, 3, 5-40, 2001.

\bibitem{almgrenciti}R. Almgren, C. Thum, E. Hauptmann, H. Li. Direct estimation of equity market impact. Risk, 18(7):58–62, 2005.


\bibitem{bargeron}
L. Bargeron, M. Kulchania, S. Thomas.  Accelerated share repurchases. Journal of Financial Economics, 101(1), 69-89, 2011.

\bibitem{cj1}
Á. Cartea, S. Jaimungal, J. Ricci. Buy low sell high: a high frequency trading perspective. SIAM Journal on Financial Mathematics, 5(1), 2014

\bibitem{cj2}
Á. Cartea, S. Jaimungal, Risk Metrics and Fine Tuning of High Frequency Trading Strategies, 25(3), Mathematical Finance, 2015.

\bibitem{chemmanur}
T. J. Chemmanur, Y. Cheng, T. T. Zhang. Why do firms undertake accelerated share repurchase programs?, 2010

\bibitem{gatheral}  J. Gatheral, No-Dynamic-Arbitrage and Market Impact, Quantitative Finance, 10(7), 749-759, 2010


\bibitem{gueant}O. Guéant, Optimal execution and block trade pricing: a general framework, Applied Mathematical Finance, 22(4),  2015

\bibitem{gueantpov}
O.~Gu{\'e}ant, Execution and block trade pricing with optimal constant rate
  of participation, {\em Journal of Mathematical Finance}, 4(4),   2014.

\bibitem{gueantpu} O. Guéant, J. Pu, Option pricing and hedging with execution costs and market impact, to appear in Mathematical Finance.

\bibitem{asr} O. Guéant, J. Pu, G. Royer, Accelerated Share Repurchase: pricing and execution strategy, International Journal of Theoretical and Applied Finance, 18(3),
  2015.

\bibitem{book} O.~Gu{\'e}ant.
\newblock The Financial Mathematics of Market Liquidity: from Optimal Execution to Market Making.
\newblock {\em CRC Press, Taylor and Francis}, 2016.

\bibitem{jaimungal}S. Jaimungal, D. Kinzebulatov, D. Rubisov, Optimal Accelerated Share Repurchase, preprint, 2013

\bibitem{lehalle1}C.-A. Lehalle, S. Laruelle, R. Burgot, S. Pelin,
M. Lasnier, Market Microstructure in Practice. World Scientific publishing,
2013.

\bibitem{lehalle2}C.-A. Lehalle, M. Lasnier, P. Bessson, H. Harti,
W. Huang, N. Joseph, L. Massoulard, What does the saw-tooth pattern
on US markets on 19 july 2012 tell us about the price formation process.
Crédit Agricole Cheuvreux Quant Note, Aug. 2012.





\bibitem{tcleland}H. E. Leland, Option pricing and replication with
transactions costs. The Journal of Finance, 40(5), 1283-1301, 1985


\bibitem{li}T. M. Li, R. Almgren, Option Hedging with Smooth Market Impact, preprint, 2014.

\bibitem{rogerssingh}L. C. Rogers, S. Singh, The cost of illiquidity
and its effects on hedging. Mathematical Finance, 20(4), 597-615,
2010.


\end{thebibliography}

}

\end{document}